# A multistep pulse compressor for 10s to 100s PW lasers

**JUN LIU**[1,2,3], **XIONG SHEN**[1,2], **SHUMAN DU**[1,2], **RUXIN LI**[1,2,4]

[1] *State Key Laboratory of High Field Laser Physics and CAS Center for Excellence in Ultra-intense Laser Science, Shanghai Institute of Optics and Fine Mechanics, Chinese Academy of Sciences, Shanghai 201800, China*
[2] *University Center of Materials Science and Optoelectronics Engineering, University of Chinese Academy of Sciences, Beijing 100049, China*
[3] *jliu@siom.ac.cn*
[4] *ruxinli@mail.siom.ac.cn*

**Abstract:** High-energy tens (10s) to hundreds (100s) petawatt (PW) lasers are key tools for exploring frontier fundamental researches such as strong-field quantum electrodynamics (QED), and the generation of positron-electron pair from vacuum. Recently, pulse compressor became the main obstacle on achieving higher peak power due to the limitation of damage threshold and size of diffraction gratings. Here, we propose a feasible multistep pulse compressor (MPC) to increase the maximum bearable input and output pulse energies through modifying their spatiotemporal properties. Typically, the new MPC including a prism pair for pre-compression, a four-grating compressor (FGC) for main compression, and a spatiotemporal focusing based self-compressor for post-compression. The prism pair can induce spatial dispersion to smooth and enlarge the laser beam, which increase the maximum input and output pulse energies. As a result, as high as 100 PW laser with single beam or more than 150 PW through combining two beams can be obtained by using MPC and current available optics. This new optical design will simplify the compressor, improve the stability, and save expensive gratings/optics simultaneously. Together with the multi-beam tiled-aperture combining method, the tiled-grating method, larger gratings, or negative chirp pulse based self-compression method, several 100s PW laser beam is expected to be obtained by using this MPC method in the future, which will further extend the ultra-intense laser physics research fields.

1. **Introduction**

Owing to the application of chirped pulse amplification (CPA) in laser research field in 1985 [1], the peak power of laser pulse had been broken down as long as 20 years plateau and gotten great progress since then. According to a recent review paper, tens of petawatts (PW) laser facilities had been built up around the world [2]. Among them, both the SULF facility in China (SULF-10 PW) and the ELI-Nuclear Physics (ELI-NP) in Romania have already achieved as high as 10 PW laser pulse recently [3,4], which is the highest peak power laser up to now. Several other 10-PW-level laser facilities are in building, such as the APPOLON in France, the upgrading Vulcan laser located at the Central Laser Facility in United Kingdom, and PEARL-10 PW at the Institute of Applied Physics of the Russian Academy of Sciences in Russia [5-7]. To achieve higher peak power and focal intensity, several tens (10s) to hundreds (100s) PW level laser facilities had been proposed all over the world, such as the XCELS-200PW (Exawatt Center for Extreme Light Studies, Russia) [7], the ELI-200PW (Extreme Light Infrastructure, Europe) [8], the OPAL-75PW (Optical Parametric Amplifier Line, USA) [9], and the SEL-100PW (Station for Extreme Light, China) [10]. Among them, the SEL-100PW facility had been started up in 2018 [10-11]. Together with the intense hard X-ray laser, this SEL-100PW laser will be used to study strong-field quantum electrodynamics (QED), vacuum birefringence, even the generation of positron-electron pair from vacuum [8-10], which can extend the basic knowledge for humankind in the physical world.

In an ultrahigh intense PW laser system, the laser pulse energy is firstly amplified by using two famous techniques, which are the CPA technique based on laser crystals and the optical parametric chirped pulse amplification (OPCPA) based on nonlinear optical crystals [12]. And then, the amplified positively chirped nanosecond pulse is compressed back to ultrashort femtosecond pulse by using a grating-based pulse compressor. However, new limitation appeared during achieve 10s to100s PW laser pulses recently. This time, the limitation did not come from pulse amplification but from the pulse compression. So far, it seemed impossible to achieve a 10s to 100s PW laser pulse directly from a single traditional grating-based pulse compressor because the damage threshold and the maximum size of diffraction gratings are not high or large enough to satisfy the requirement currently [13].

To obtain 10s to 100s PW laser pulse, the multi-beam tiled-aperture combing was the main scheme which were proposed in the XCELS-200PW, the ELI-200PW, and the SEL-100PW facilities. This tiled-aperture beam combing method was proposed in 2006 at first [14]. Compared with the tiled-grating method [15], the tiled-aperture beam combing shows the advantage that fewer parameters need to be controlled [14, 16]. By using small-sized beams with femtosecond pulse duration, many theoretical and experimental studies have been done to prove the feasibility of the tiled-aperture beam combining method [17-19]. Furthermore, with relatively large beam size, 1.15-PW PETAL was obtained by using the tiled-aperture beam combining method in 2017 [20]. However, this beam combining method is still very complicated which is very sensitive to the differences of optical delay, pointing stability, wavefront, and dispersion among every beams [16-19]. In the optical design of the proposed OPAL-75PW facility, a potential relative large size of grating with higher damage threshold and bigger size was expected to achieve 75 PW with single beam. However, there is no report that this kind of grating is available right now. Besides beam combing methods, 10s to 100s PW with relative shorter pulse duration were also proposed recently by using wide-angle non-collinear OPCPA [21], post compression based on self-phase modulation in thin plates [21-22].

In this paper, we propose a feasible new multistep pulse compressor (MPC) for achieving high-energy 100 PW laser pulse without neither beam splitting nor beam combining, but using a single beam and multiple compression steps. By modifying the spatiotemporal properties of the input and output beams, we can increase the maximum bearable pulse energies in a four-grating compressor (FGC). This new MPC typically including a prism pair for pre-compression, a FGC for main compression, and a spatiotemporal focusing based self-compressor for post-compression. The prism pair used in the pre-compressor induces spatial dispersion which simultaneously smooths and enlarges the input laser beam. Then, it can increase the input pulse energy. Together with the spatiotemporal focusing effect in the post-compressor, it is expected to obtain 10s to 100s PW laser pulse by using this new method with current available gratings. Moreover, by using the self-compression effect in thin plates with negatively chirped pulse input, even several 100s PW laser pulse with shorter pulse duration is expected to be obtained in the future. This new optical design will simplify the compressor, improve the stability, and save expensive gratings/optics simultaneously because a single FGC is used. Furthermore, owing to the beam smoothing effect of the pre-compressor, the MPC method can be extended to all traditional PW laser systems, which improves the operating safety of PW lasers because it reduces the damage risk from the amplified laser beam with strong spatial intensity modulation or hot spots.

## 2. Principle and scheme of multi-step pulse compressor

Recently, a novel "in-house beam-splitting compressor" design has been proposed to reduce the beam combing challenge for 10s to 100s PW lasers [23]. In comparison to the XCELS-200PW, the ELI-200PW, and the former SEL-100PW designs, in which the beam splitters were located before the amplifier or before the compressor, respectively, the beam splitter in this new "in-house beam-splitting compressor" is moved backward into the compressor, which is located between the second grating and the third grating of a typical FGC. As a result, the novel in-house design saves at least four expensive meter size gratings and some related optics because it shares the first pair of gratings. Even though the "in-house beam-splitting compressor" has greatly reduced the complex, difficulty, and cost on achieving 10s to 100s PW laser pulse, the tiled-aperture combining of four beams is needed. And it seems that the beam splitter cannot be moved backward further. Is there any method or design that can further simplify the compressor? Here we propose a novel "multistep pulse compressor" which is able to achieve 10s to 100s PW laser pulse with even single beam in the whole compressor. The original idea or principle of the proposed MPC is shown in Figure 1.

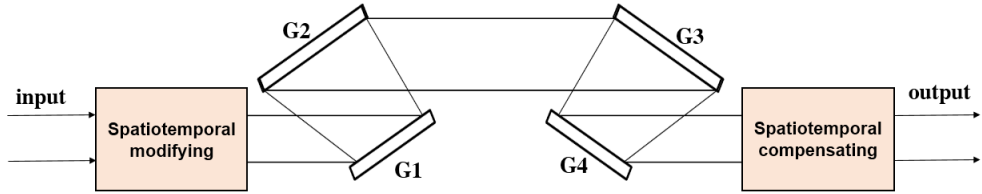

Fig.1. Principle of MPC. Spatiotemporal modifying and spatiotemporal compensating modules are added before and after the main FGC. G1-G4 are diffraction gratings for a typical FGC.

Usually, in a high-energy PW laser system, typical FGC is used for the final pulse compression, where the first grating bears the highest input pulse energy and the last or fourth grating stands the shortest output pulse duration. The damage thresholds of the first grating and the last grating limit the maximum input and output pulse energies, which are hard to be improved by the grating manufacture currently. Besides the gratings, the damage threshold is also limited by the spatiotemporal properties of the input and output laser beams in a FGC. As a result, we can find resolutions from shaping or modifying the spatiotemporal properties of the input and output laser beams. In the previous "in-house beam-splitting compressor", the property that laser-induced damage threshold of gratings decreases with the shortening of pulse duration was passively used [23]. Here, through actively modifying both the temporal and the spatial properties of the input and output laser beams on the surface of the first and the last gratings, respectively, it is possible to improve the maximum input and output pulse energy in the FGC. To realize this goal, a pre-compressor and a post-compressor are used to modify and compensate the spatiotemporal properties of the input and the output laser beams, respectively. Then, we named this new compressor as "multistep pulse compressor" because it divides traditional simple single FGC pulse compression process into several steps or stages. Typically, there are three steps or stages including pre-compressor, main compressor, and post-compressor. The pre-compressor is used to modify input laser spatiotemporal parameters so as to increase the input pulse energy. On the same way, the output laser can own modified spatiotemporal parameters to satisfy a high output pulse energy, which is then compensated by using a post-compressor.

The basic optical diagrammatic sketch of the proposed MPC to achieve 10s to 100s PW laser is shown in Figure 2. A prism pair based compressor is used to induce suitable spatial dispersion along one axis (hereafter named X axis) perpendicular to the direction of propagation of the input laser beam in the optical propagation plane. The induced spatial dispersion will smooth the laser intensity in the spatial domain and enlarge the laser beam simultaneously. Then, it will increase the maximum input pulse energy for the main FGC compressor. Note that another prism pair can be used along the orthogonal axis (hereafter named Y axis) to further smooth the laser beam, which is not shown in Figure 2. The main compressor used for PW laser system is a FGC because it can induce enough negative chirp to compensate most of the positive chirp from the front stretcher. With a symmetric FGC, the output laser beam will also own intensity smoothed spatial profile. As a result, it will increase the maximum output pulse energy through increasing the maximum tolerable pulse energy on the fourth grating. However, the spatiotemporal properties of the laser pulse come out from the main compressor are modified. Then, after the main compressor, the spatial dispersion has to be further compensated to achieve ultrashort femtosecond laser pulse at the focal point. Here, the spatiotemporal focusing effect is used to compensate the spatial dispersion at the focal point automatically [24-25]. Furthermore, with the technique of self-compression of negative chirped pulses using thin plates [26], pulse much shorter than that of the Fourier transform limitation (FTL) duration of the input pulse is expected to be obtained, which will further increase the peak power and the focal intensity. Note that the FGC used here may owe asymmetric distances between the first pair of gratings (G1 and G2) and the second pair of gratings (G3 and G4), which can be called asymmetric FGC. With an asymmetric FGC, it is possible to generate a spatial dispersion free output laser beam but with intensity modulated beam profile, which will be discussed in the following sections.

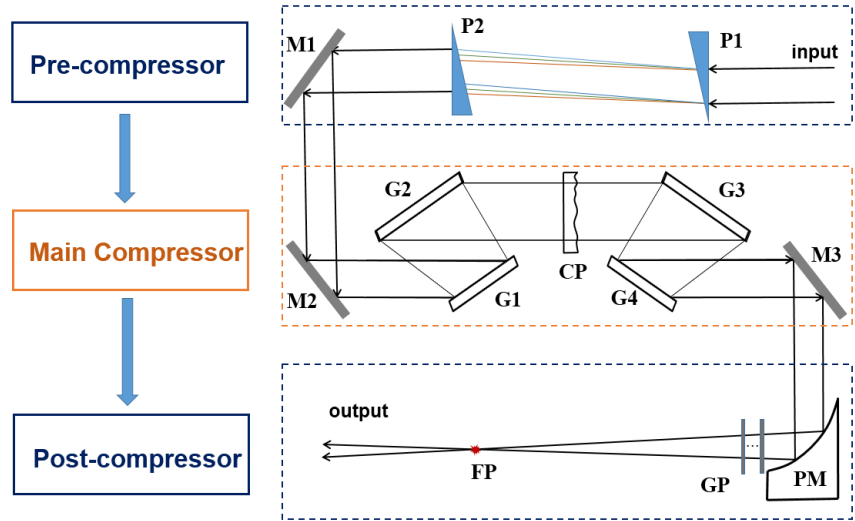

Fig.2. The basic scheme of MPC including pre-compressor, main compressor, and post-compressor. P1, P2: Prism, M1-M3: reflective mirrors, G1-G4: diffraction gratings, CP: compensating plate, GP: thin glass plates, PM: parabolic mirror, FP: focal point.

### 3. Prism pairs based compressor for pre-compression

The pre-compressor is used to modify the spatiotemporal properties of the input laser beam on the first grating of the main FGC. As we know the spatial intensity modulation of the amplified laser beam is relatively high in PW laser system, which is mainly due to the high spatial intensity modulation of the pump beam usually. Moreover, some hot spots may appear on the surface of the first diffraction grating in traditional setup due to diffraction by dust or defect on optics before the FGC. When considering these negative influences, in a typical engineering design of FGC, the maximum input pulse energy has to be decreased by even two times to avoid the damage of the first or last grating of the main FGC.

Therefore, if we can reduce the spatial intensity modulation of the input laser beam, we can increase the maximum input pulse energy for the main FGC. Inspired by our previous "in-house beam-splitting compressor" work, in which the beam profile after the first pair of gratings is excellently smoothed in comparison to that of the input laser beam with strong spatial intensity modulation [23]. The beam smoothing effect is owing to the spatial dispersion of the laser beam by the grating pair. Usually, strong spatial intensity modulation appeared at relative high spatial frequencies which is the main problem, while the spatial intensity modulation appeared at the low spatial frequency is small. As a result, there is no need of very large spatial dispersion to smooth the laser beam in a wide spatial range. Furthermore, if considering the loss of diffraction gratings, we propose prism pairs for the generating suitable spatial dispersion so as to smooth the laser beam. When considering a very large size of amplified laser beam in PW laser, the prism is better to own a relative small apex angle so as to reduce the thickness of the prism. Of course, transmitted gratings can also be used to induce the spatial dispersion if there are available ones.

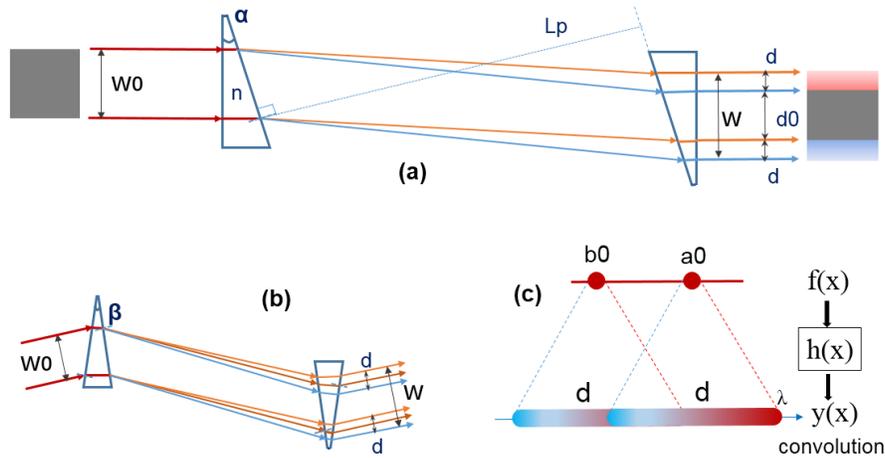

Fig.3. (a). The optical setup of (a) right angle prism pair and (b) isosceles prism pair for pre-compression. $W0$: input laser beam width, $W$: output laser beam width, $d$: spatial dispersion width, $Lp$: the perpendicular distance between the two prisms, $\alpha$ and $\beta$ are the apex angles of the right angle prism and isosceles prism, respectively. (c) Principle of beam smoothing using angular/spatial dispersion. a0 and b0: two hot spots with high intensity and full spectral bandwidth, $d$: extended length of a0 or b0 after the prism pair.

Two kinds of configurations were considered for single prism pair: 1) right angle prism pair and 2) isosceles prism pair. The optical setup is shown in Figure 3(a) and Figure 3(b), respectively. As for the right angle prism pair configuration shown in Figure 3(a), only the hypotenuse surface will introduce spatial dispersion to the input beam with a zero incident angle on the right-angle side, where $Lp$ is the distance between the two prism, $\alpha$ is the apex angle, $d$ is the induced spatial dispersion width, $W0$ and $W$ are the input and output beam widths along the X axis, respectively. Then, $d$ is linearly related to $Lp$. For the isosceles prism pair, both hypotenuse will induce angular dispersion to the input beam with a nonzero incident angle, as shown in Figure 3(b). As a result, to generate same amount of spatial dispersion, the distance of the prism pair for the isosceles prism pair will be much shorter than that of the right angle prism pair. In a word, the right angle prism owns thin thickness, while the using of isosceles prism with larger apex angle will shorten the prism distance. Here, only the right angle prism pair is discussed which is simple and easy to understand the proposed method.

Figure 3(c) shows the basic principle of beam smoothing by using angular dispersion induced spatial dispersion, where a0 and b0 are two hot spots with high intensity on the beam. The beam radius of both a0 and b0 are assumed to be $r_0$. After the prism pair, both spots have been extended into lines with a length of d and a width of $2\times r_0$. As a result, the peak intensity is decreased by about $2\times d\times r_0/(\pi\times r_0^2)= 2\times d/(\pi\times r_0)$ times for these two spots. After summing up all the corresponding spatial and spectral intensity values at every point on the extended beam line, the intensity of the output laser beam is smoothed. The beam smoothing effect is related to the ratio of $d/r_0$. It means larger spatial dispersion width is help to smooth the laser beam. It is easier to smooth the intensity modulation with high spatial frequency owing to $d/r_0>>1$. This angular dispersion induced spatial dispersion process can also be explained by using a convolution, as shown in Figure 3(c) and the following expression.

$$y(x) = \sum_{n=-\infty}^{\infty} f(n)h(x-n) = f(x) \otimes h(x)$$

Where $f(x)$ is the laser intensity distribution function along the X axis, $h(x)$ function is the spectral intensity profile projecting on the X axis, $y(x)$ is the obtained smoothed profile. As we know that the convolution operator owns the ability of smoothing figures or signals.

Some calculations and simulations were performed to analysis the induced spatial dispersion and beam smoothing by the right angle prism pair. Here, the input laser beam is assumed owning 10$^{th}$ order super-Gaussian spatial profile with a spatial intensity modulation ratio (peak over average) of 2. The beam size is 370mm×370mm. A random spatial intensity modulation was lead along the beam with a spatial frequency of about 1 mm$^{-1}$. The spectrum is also a 10$^{th}$ order super-Gaussian profile with smoothing intensity and a spectral range from 825nm to 1025nm. The right angle prism is assumed the fused silica glass prism. The apex angle α at 15º, 25º, and 30º three different angles are discussed. Figure 4(a) shows the spatial intensity modulation ratio decreasing as the increasing of spatial dispersion width $d$. It can be seen the spatial intensity modulation ratio is decreased rapidly from above 2.0 to near 1.2. The beam smoothing effect is very efficient. However, it takes relative long distance to further smooth the modulation ratio from 1.2 to 1.1. As expected, it takes much shorter distance to realize same modulation ratio and spatial dispersion width $d$ for the prism pair with 30º apex angle over that of the 15º one. The decreasing of spatial intensity modulation ratio show almost the overlapped curves at three different angles, which means the spatial intensity modulation ratio of the smoothed beam is related to the spatial dispersion width $d$ tightly. For the same spatial dispersion width of about 60 mm at three different apex angles, the spatial intensity modulation of the beam has all been well smoothed to about 1.1. Figure 4(b-g) show both the one dimensional and the two dimensional beam profiles of the input laser beam and those of the smoothed output laser beam. In the case, the smoothed beam own about 60 mm spatial dispersion width along the X axis. Obviously, the laser beam is well smoothed on both axes with the spatial intensity modulation of about 1.1.

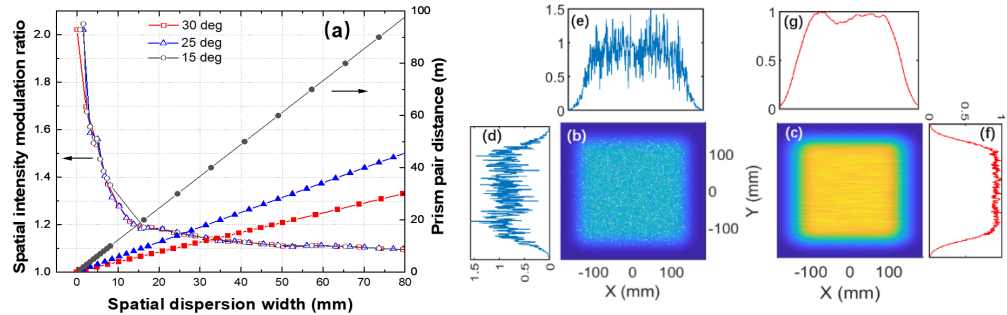

Fig.4. (a) The decreasing of the spatial intensity modulation ratio and the increasing of the prism pair distance according to the increasing of the induced spatial dispersion width at 30º, 25º, and 15º three different apex angles of the prism. The two dimensional profiles of (b) the input laser beam and (c) the smoothed laser beam. The one dimensional spatial intensity profiles of (d-e) the input beam and (f-g) the smoothed beam at both X and Y axes.

When the input laser spectrum also owns intensity modulation, it is found that the beam smoothing effect is still highly active. The spatial intensity modulation ratio decreasing is almost the same to that without spectral intensity modulation one. This is owing to the convolution process between the spatial intensity profile and the spectral intensity profile. Owing to the big size of laser beam and relative thick prism, the induced spectral dispersion by the prism pair can be positive or negative chirp at different distances. It means that the prism pair with small apex angle can induce large spatial dispersion with relatively small spectral dispersion. Anyway, the induced small spectral dispersion can be easily compensated by using the FGC.

From above simulation results, it can reduce the spatial intensity modulation effectively through the spatial dispersion along the X axis of the amplified laser beam by using one prism pair. Actually, it is possible to introduce another spatial dispersion along the Y axis to the laser beam when another prism pair placed perpendicular to the first one. From the calculation results above, the spatial intensity modulation decreased rapidly at the beginning. It means a relative small spatial dispersion width about 10 mm induced along the Y axis can further smooth the laser beam rapidly to a smaller value, which means it is feasible to achieve a smoothed beam with 1.1 spatial intensity modulation. The spatial dispersions along both X and Y axes can be compensated by using spatiotemporal focusing effect in the post-compressor. As for the spectral dispersion induced along the Y axis, a thick glass plate can be used to completely compensate the induced negative chirp. If the induced spectral dispersion along Y axis is positive chirp, the FGC can compensate its dispersion. Detail calculation and simulation results about the spatial dispersion, the spectral dispersion and the beam smoothing of the pre-compressor with prism pair will be shown in a separated paper.

## 4. A four-grating compressor for main compression

Owing to the capability of inducing enough negative chirp to compensate the large amount of positive dispersion induced by the grating-based stretcher, FGC is automatically used as the main compressor in the MPC. The incident angle and diffraction angle on the first grating G1 is $\theta$ and $\gamma(\lambda)$ at $\lambda$ wavelength, respectively. From the dispersive expressions of a parallel grating pair, the spectral dispersion of a typical FGC is related to the perpendicular distances of both the first pair of gratings (G1 and G2) and that of the second pair of gratings (G3 and G4), which are $L1$ and $L2$, respectively [27].

Except for beam smoothing, the laser beam after the prism pair is also enlarged due to spatial dispersion simultaneously. The spatial dispersion values between the prism pair in pre-compressor and the grating pair in FGC can be set to same or opposite. If their spatial dispersions are the same, it will induce serious spectral cutting and energy loss to the input laser beam. Fortunately, when the spatial dispersion of the input beam is set opposite to the induced spatial dispersion by the FGC, the input laser beam is possible not be diffracted out of the second grating G2 if the induced spatial dispersion from pre-compressor is smaller than that induced by the grating pair. That is $D1=D2 \leq D$ shown in Figure 5, which is very easy to be achieved. Meanwhile, the positive or negative value of the input spatial dispersion can be conveniently tuned by overturning the direction of the two prisms in the pre-compressor. Then, we only discuss in detail the condition that opposite spatial dispersions between the pre-compressor and the FGC are applied.

Figure 5(a) shows the optical configuration in the condition of the maximum input beam size without beam cutting or spectral shearing, in which the input beam owns opposite spatial dispersion to that induced by the grating pair. In the figure, $D$ is the maximum extended length of a full spectral bandwidth laser on the G2, where $D=d/\cos\theta$. As we can see that the maximum input beam size on the first grating G1 can even larger than the beam on the G2 by $D$ along the grating direction. That is to say, besides the smoothed center dark grey part ($D0$ or $d0$) with full spectral bandwidth can increase the input pulse energy, the red region with longer wavelength and the blue region with shorter wavelength ($D$ or $d$) are absolutely complimentary parts that can also increase the input pulse energy. If the maximum energy of the smoothed dark grey region is set at a safe 1.33 times lower than the threshold, in comparison to previous typical 2 times, the maximum input pulse energy will be improved by about $2/1.33 \approx 1.50$ times. The complimentary red and blue regions with spatial dispersion increase the input pulse energy by about $(d0+d)/d0$ times. Therefore, this configuration can increase the input pulse energy greatly.

The far edge of both blue and red regions own single frequency in principle, around where the pulse energy is very weak. It means that it is possible further increase the input pulse energy by cutting this weak light owning narrow bandwidth. This step can be done together with enlarging dark grey region and introducing larger spatial dispersion which will further smooth the laser beam. Therefore, the maximum input pulse energy will be further increased. Note that there is energy loss for the input laser beam in the case.

Since the maximum size of a usable grating is fixed in reality, as a result, we actually can only obtain the same maximum size of gratings for both grating G1 and grating G2 to achieve the maximum input and output pulse energies. Moreover, the beam on G2 is also well smoothed in the case. Then, the input beam with spatial dispersion from the pre-compressor can be designed exactly fill in the grating G1 with the maximum size, as shown in Figure 5(b), where $D1+D2 \leq D$, $D0=d0/\cos\theta$. Typically, $D1=D2 \leq D/2 < D$, $d1=d2 \leq D \times \cos\theta/2$. According to the principle of the FGC with grating pairs [27], $L0$, which is the perpendicular distance between G1 and G2 of a typical FGC, can be expressed as:

$$L0 = \frac{cT_0 d^2}{2\lambda_0(\lambda_l - \lambda_s)}\left[1 - \left(\frac{\lambda_0}{d} - \sin\theta\right)^2\right]^{\frac{3}{2}} \quad (1)$$

Where $c$ is the light speed in vacuum, $T_0$ is the chirped pulse duration of the input beam for the FGC, $d$ is the line density of the grating, $\lambda_0$ is the central wavelength of the laser, $\lambda_l$ and $\lambda_s$ are the longest and shortest wavelengths of the input laser beam, respectively. Then, g, the gap between the far edge light with the longest wavelength and the edge of G1, and $D$ can be expressed as:

$$D = [\tan\gamma(\lambda_l) - \tan\gamma(\lambda_s)]L0 \quad (2)$$

$$g = \{[\tan\theta - \tan\gamma(\lambda_l)]L0 - D0 - D2\}\cos\theta \quad (3)$$

From the above expressions, the grating pair distance $L0$, the incident angle $\theta$, $D0$ and $D2$ should be considered together so as to let the gap $g > 0$. It is possible further increase the input and output pulse energy by cutting weak lights owning narrow bandwidth on the far edge of both blue and red regions in reality design.

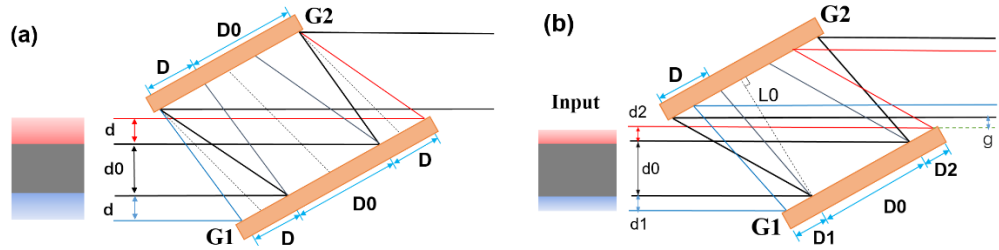

Fig.5. The optical configuration (a) of the first pair of gratings for the maximum input beam size without beam cutting or spectral shearing; (b) of the first pair of gratings for the maximum available gratings without beam cutting or spectral shearing. Where, $L0$: the perpendicular distance between G1 and G2, g: the gap between G1 edge and the output beam edge, D: the maximum extended length of a full spectral bandwidth laser on G2, $dn=Dn \times \cos\theta$, $n=0, 1, 2$.

As for a FGC, the induced spectral dispersion are positively related to the sum distances of the two parallel grating pairs, that is $L1+L2$. For a basic typical FGC, the two prism pairs are symmetric with $L1=L2$, as shown in Figure 6(a). There are other two asymmetric arrangements, which are $L1>L2$ (shown in Figure 6(b)) and $L1<L2$. In asymmetric cases, the spectral dispersions can also be completely compensated while it will introduce spatial chirp to the output laser beam. Then, two kinds of optical configurations are considered here. They are the symmetric arrangement with smoothed output laser beam and the asymmetric arrangement with spatial dispersion free output laser beam.

For a typical symmetric FGC, the perpendicular distance between the first pair of gratings G1 and G2, is equal to that of the second pair of gratings G3 and G4. These two parallel grating pairs are mirroring each other. Here, a compensating plate (CP) is used to compensate the diffraction wavefront errors of grating G2 and grating G3 [23]. The CP can also be used to compensate the negative spectral dispersion induced by the prism pair in the pre-compressor. The detail optical configuration of the symmetric FGC located after the pre-compressor is shown in Figure 6(a), where $D1=D2 \leqslant D/2$. For simplification, the input laser beam only show spatial dispersion along the X axis, where the dark grey region owns full spectral bandwidth, the red and blue region own partial spectrum together with spatial dispersion. It is very clear that the output owns same smoothed beam with spatial dispersion which will increase the maximum tolerant output pulse energy. With an opposite spatial dispersion setting to the FGC, the red and blue regions with spatial dispersion will be diffracted inside onto the second grating G2. As a result, there are no loss of pulse energy or cutting of spectrum during the whole FGC process.

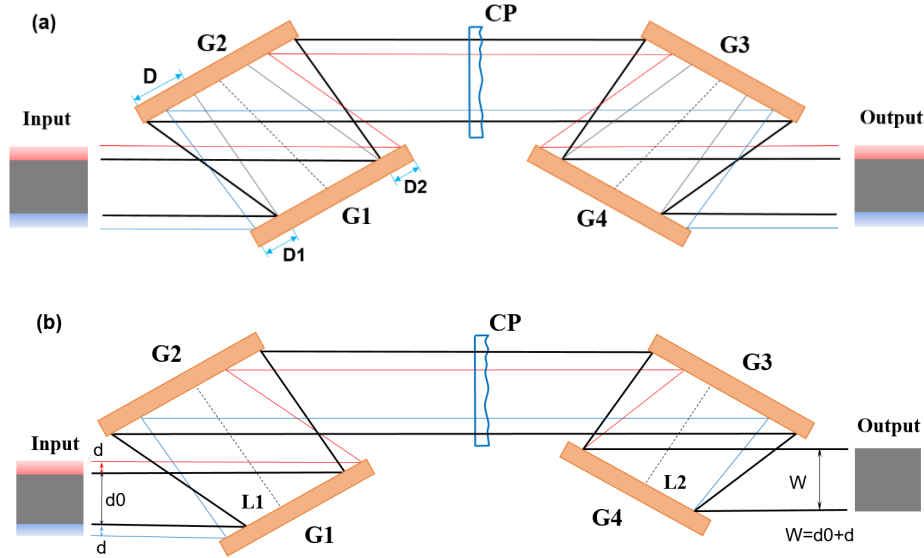

Fig.6. The optical configurations of (a) the symmetric FGC and (b) the asymmetric FGC. G1-G4: diffraction gratings, CP: compensating plate, $L1$, $L2$: perpendicular distances of the first and the second grating pairs, respectively. D: the maximum extended length of a full spectral bandwidth laser on G2, $dn=Dn\times\cos\theta$, $n$=0, 1, 2. The input beam own opposite spatial dispersion with the FGC. The dark grey region ($d0$) own full bandwidth spectrum, while the red and blue regions (d) own partial spectrum and spatial dispersion.

The symmetric arrangement induces smoothed output laser beam, while the asymmetric arrangement can induce spatial dispersion free output laser beam along the X axis. The optical configuration is shown in Figure 6(b), where $L1>L2$. With suitable different distance between $L1$ and $L2$, the spatial dispersion of the input smoothed laser beam can be well compensated. The output laser beam size along the X axis is $W=d0+d$. Note that the output beam will own spatial intensity modulation the same as the input laser beam of the pre-compressor if there is no spatial dispersion along the Y axis. In another word, the output beam can also be a smoothed laser beam if the input laser beam own spatial dispersions on both X and Y axes.

The pre-compressor based on prism pair has already induced certain temporal chirp and spatial dispersion to the laser beam, while the CP will induce positive spectral dispersion. Actually, the spectral dispersion induced by the prism pair is small. Then, the influence to FGC in the spatial domain can be neglected in comparison to the large laser beam size. To achieve a laser pulse of which the spectral dispersion is completely compensated, the distances of the two parallel grating pairs should be tuned slightly. The additional spectral dispersion induced by FGC should be equal to the total spectral dispersion induced by the pre-compressor and CP. Then, the output own FTL femtosecond pulse duration for the central dark grey region, while the blue and red regions own different spectral bandwidths and different pulse durations. At the same time, the output laser beam is the same as the input beam which is well smoothed in spatial domain. It needs the spatiotemporal focusing effect to achieve compressed pulse at the focal point for both blue and red regions which will be discussed in the next section.

Besides the spatial intensity modifying, the temporal profile can also be modified to improve the output pulse energy of the main FGC. Note that the damage threshold of the diffraction grating is related to the pulse duration on the surface [13]. This property had been successfully used in the "in-house beam-splitting pulse compressor" design [23]. From experimental data and tests from commercial products, the highest damage threshold ratio of energy density (mJ/cm$^2$) for nanosecond, picosecond and femtosecond laser pulses is about 600:380:229 =2.67:1.66:1 for gold-coated diffraction gratings at 800 nm central wavelength. Then, if we can let the output pulse duration at the surface of the last grating of the FGC operate at picosecond level, we can further improve the output pulse energy by about 1.6 times. However, the spectral dispersion need to be well compensated which will be discussed in the next section.

## 5. A self-compressor for post-compression

Since the spatiotemporal properties of the output laser beam from the FGC are shaped to increase the maximum sustainable output pulse energy of the main compressor, there has to be a further step to compensate the spatiotemporal properties so as to achieve the expected extremely strong focal intensity. From the description in above several sections, the output laser beam from the FGC owns spatial dispersion to smooth and enlarge the laser beam, or owns temporal chirp, or even both. Then, a post-compressor is proposed to compensate both of the spectral dispersion and the spatial dispersion of the output laser.

In the spatial domain, to eliminate the influence of the spatial dispersion along both of the X and Y axes, the spatiotemporal focusing technique is used, which has been successfully used in two-photon microscopy and ultrafast micro-machining applications [24-25]. As for the output laser beam, the pulse duration on four edges is relative wide due to the narrow spectrum. With the spatiotemporal focusing effect, the four edges of the output beam with spatial dispersion will be compressed automatically in time domain at the focal point. The same as spatiotemporal focusing used in two-photon microscopy to improve the spatial resolution, this spatiotemporal focusing configuration will induce almost diffraction limited focusing and FTL pulse duration at the focal point. Special spatiotemporal light field appears around the focal point which may be useful in high field laser physics researches.

Figure 7 shows the calculated spatiotemporal properties at the focal plane with different spatial dispersion on the beam. In the calculation, a 370×370 mm$^2$ laser beam with 10$^{th}$ order super-Gaussian spatial profile is focused using a parabolic mirror with 3 meters focal length. The FTL pulse duration of the laser pulse is 14.5 fs, which is centered at 925 nm with a 200 nm full bandwidth. Figure 7(a-c) show spatiotemporal properties at the focal point when there is no spatial dispersion ($\Delta x$=0 mm, $\Delta y$=0 mm). Figure 7(d-f) show spatiotemporal properties at the focal point when there is spatial dispersed along the X axis only. The induced spatial dispersion width $d$ is set to 100 mm ($\Delta x$=100 mm, $\Delta y$=0 mm). We can see that even though the spatial dispersion along the X axis induces a pulse front tilt in the Y plane (Fig. 7(e)). In the time domain, the induced 100 mm spatial dispersion along the X axis slightly broaden the pulse duration at the focal point by about 0.5 fs, as shown in Figure 7(i), where the red curve shows a 14.5 fs FTL pulse duration without spatial dispersion and the blue line shows about 15 fs pulse duration with spatial dispersion. In the spatial domain, the focal spots are almost the same for laser beam without and with the 100 mm spatial dispersion along the X axis, as shown in Figure 7(c) and 7(f), respectively. As a result, both the focal diameter and the pulse duration at the focal point are barely affected by the induced spatial dispersion. It means that the spatiotemporal focusing technique is a very simple and effective method can be used to self-compensate the induced spatial dispersion during the post compression process.

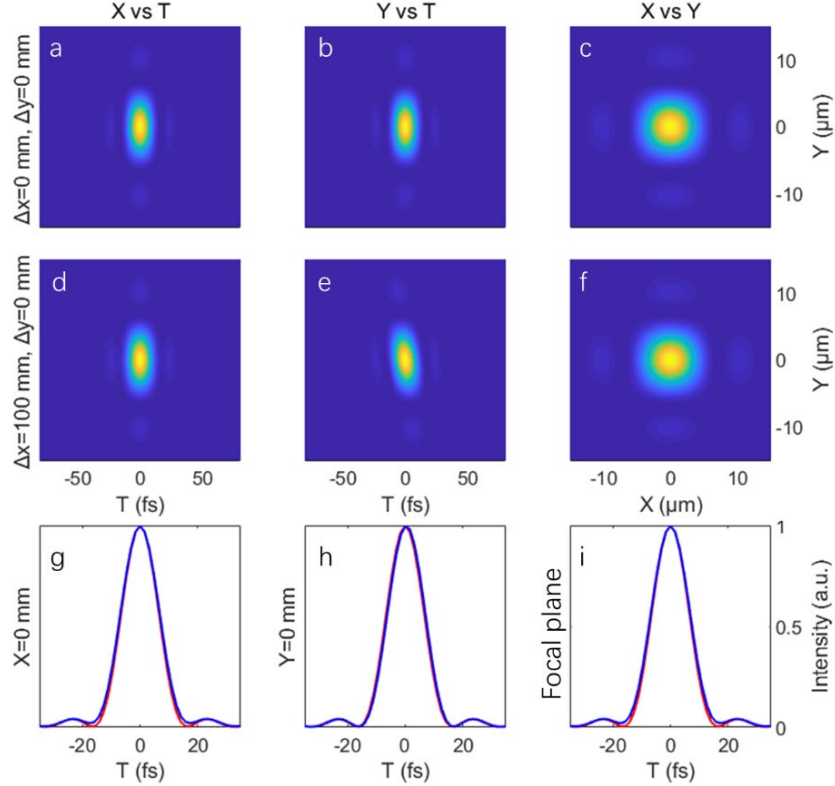

Fig.7. The spatiotemporal properties at the focal plane with and without spatial dispersion on the laser beam. Spatiotemporal properties along X axis or Y axis, and the integrated spatial profiles at the focal plane in the case of (a-c) with no spatial dispersion or (d-f) with 100 mm spatial dispersion along the X axis, respectively. (g-h): Temporal profiles when (g) x=0 mm in X plane, (h) y=0 mm in Y plane, and (i) the integrated temporal profiles in the focal plane. Red curves: without spatial dispersion, blue lines: with 100 mm spatial dispersion along the X axis.

In the time domain, for a positively chirped picosecond pulse, chirped mirrors can be used to compensate the dispersion in principle. Currently, this kind of chirped mirror with such big size is void. Furthermore, nonlinear effect will appear in the coating layers of the chirped mirror or even damage the mirror. Normally, negatively chirped picosecond laser pulse can be easily compensated by simply using a glass plate in principle. However, it is not easy to further magnify the meter-sized output laser beam. Then, the output intensity after FGC is very high which will induce large nonlinear effect or B integral in the glass plate. For a negatively chirped laser pulse, the nonlinear effect induced B integral will narrow the laser spectrum. Therefore, it needs new idea to compensate this negative chirp besides using chirped mirrors.

Fortunately, it had been proved experimentally and theoretically that the negatively chirped laser pulse can be reshaped and self-compressed in a piece of glass plate [26]. The negatively chirped input pulse based self-compression will avoid the using of large chirped mirrors for dispersive compensation in comparison to other recent works [21-22]. During the self-compression process, as high as TW/cm² laser intensity appear in the thin glass plate which may introduce small-scale self-focusing effect (SSSF) and leads hot spots to the laser beam. This hot spots may damage the optics after the thin glass plate. To avoid the damage of meter-sized parabolic mirror, the thin glass plate for self-compression is located after the parabolic mirror. Since there is no optics after the thin glass plate, SSSF in the thin glass plates and the hot spots will not induce damage to the PW laser system. Moreover, the sphere aberration induced by even a 5 mm thick thin glass plate can be neglected for a parabolic mirror with an F number of 3. Actually, the induce sphere aberration can also be compensated by using a deformable mirror in the system.

Furthermore, as for the central smoothed output laser beam, SSSF can be well restrained or eliminated in comparison to previous works [21-22]. Several glass plates groups with thin thickness substitutes for a thick glass plate can be used to further avoid the SSSF effect [22]. Then, most of the central dark grey part of the laser beam owning full spectral bandwidth can be well compressed. It is expected that the 15 fs FTL pulse duration is possible to be compressed to sub-10 fs deduced from previous works [21-22, 26], which means hundreds PW laser pulse with few-cycle pulse duration can be obtained with this novel MPC. As for the laser on the four edges with spatial dispersion, the SPM effect is not uniform for these parts. Then, the compressed pulse may not as short as the center laser. Both numerical simulation and experiment on self-compression in a thin glass plate using the negatively chirped pulse together with smoothed spatial profile will be done in the future.

## 6. Proposed feasible designs for 100s PW laser

Here, we propose a feasible design for 100s PW laser system based on a single MPC with currently available diffraction gratings. The optical diagrammatic sketch of the whole MPC is shown in Figure 8.

The original 10$^{th}$ order super-Gaussian 370×370 mm² laser with about 4 ns pulse duration is used as the input beam. The FTL pulse duration of the input laser pulse is 14.5 fs, which is centered at 925 nm with a 200 nm full spectral bandwidth. The input laser beam is perpendicularly guided into a fused silica prism pair with 30º apex angle, which induces about 60 mm spatial dispersion along the X axis. Then, the laser beam B2 shows a size of 430×370 mm². Note that a reflective deformable mirror (DM) is used to precisely compensate the spatiotemporal phase error induced not only by optics after the prism pair but also the spatiotemporal aberration in both the stretcher and the FGC. After the DM, an image transfer system based beam expander is used to enlarge the laser beam to 700 mm on both X and Y axes. In B3 and B4, the 500×700 mm² dark grey region own full spectral bandwidth, while the blue and the red regions with 100×700 mm² each have spatial dispersion. As a result, the effective beam size along the X axis is 500+100=600 mm. The line density of the gold coated grating is 1400 lines/mm. The input pulse energy to the FGC is set to 2300 J. With an incident angle of 61º on G1, the corresponding energy intensity on the G1 is about 266 mJ/cm², which is almost 2.26 times lower than the grating damage threshold for nanosecond pulse. Assuming a feasible total compression efficiency of 66%, the output pulse energy after the FGC should be about 1510 J which corresponding to about 100 PW if compressed to about 15 fs. The pulse energy intensity on the last grating is about 174 mJ/cm², which is about 1.31 times lower than the grating damage threshold for femtosecond pulse. It means 100 PW laser can be obtained with this single MPC since the laser spatial intensity modulation is smoothed to 1.1 by the pre-compressor. After the main compressor, the spectral dispersion compensated laser pulse with 700×700 mm² beam size is focused by using a parabolic mirror. At last, we can achieve a completely compressed laser on both spatial domain and temporal domain at the focal point, which corresponding to an extremely high focal intensity. The designed input and output laser parameters are shown in table 1.

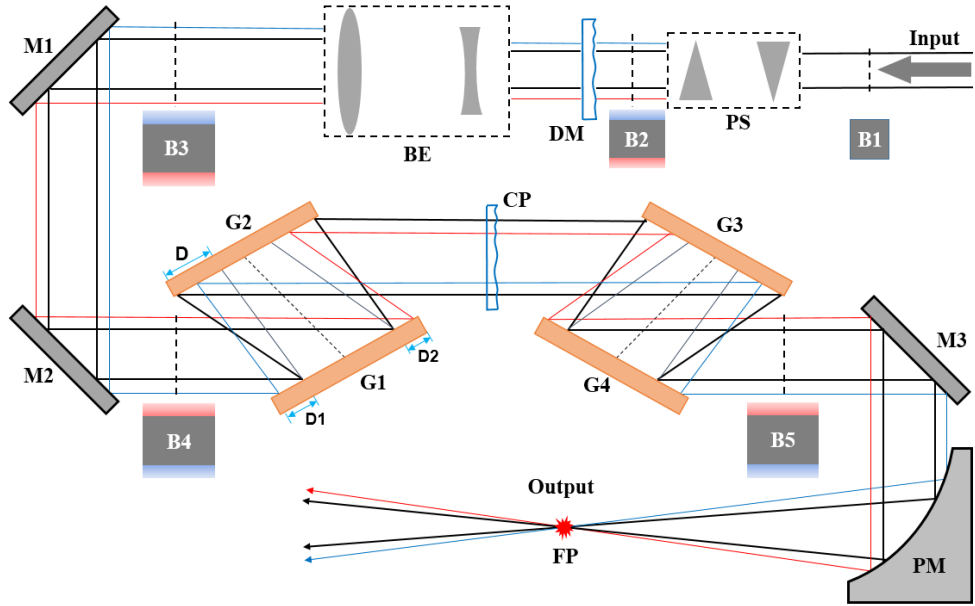

Fig.8 The optical diagrammatic sketch of the proposed single MPC for 100 PW laser. B1-5, beam profiles at five different positions indicated by the dashed lines. PS, prism pair system inducing spatial dispersion. DM, deformable mirrors for precisely spatial-spectral phase compensating besides CP. BE, beam expander. M1-3, reflective mirrors. G1-4, diffraction gratings. D, the maximum extended length of a full bandwidth laser on G2. D1-2, the extended length of blue and red regions with spatial dispersion on G1, respectively. CP, compensating plate. PM, parabolic mirror. FP, focal point.

Table 1. Designed input and output laser parameters

| | |
|---|---|
| Beam size after amplifier (full width) | 370 mm×370 mm |
| Beam intensity profile | 10$^{th}$ order super-Gaussian |
| FTL pulse duration | 14.5 fs |
| Center wavelength/full spectral bandwidth | 925 nm/200 nm |
| Spatial dispersion width $d$ | ~60 mm |
| Prism: apex angle /material | 30°/ Fused silica |
| Beam size after PS | 430 mm×370 mm |
| Beam size after BE | 700 mm×700 mm |
| The heights of blue/dark grey/red regions in B4 | 100 mm/500 mm/100 mm |
| Effective beam size after BE | 600 mm×700 mm |
| Pulse energy and energy intensity on G1 | 2300 J, ~266mJ/cm$^2$ |
| Total grating diffraction efficiency | 66% |
| Incidence angle on gratings | 61° |
| Groove density of gratings | 1400 lines/mm |
| Distances between grating pairs | ~1.24 m |
| Pulse energy and energy intensity on G4 | 1500 J, ~174 mJ/cm$^2$ |
| Pulse duration at focal point | 15 fs |

When considering the pulse energy stability and diffraction efficiency decreasing of the grating as time going by, conservative designs for long time running safety can also be proposed for 100 PW. This design is based on the tiled-aperture beam combing of two beams, which is still the simplest setup in comparison to previous proposals [7-10, 23].

In the system, the laser beam can be split using the in-house beam-splitting setup. The combining of both the MPC method and the "in-house beam-splitting pulse compressor" is stable and easy to achieve tiled-aperture beam combining. In this way, it is possible to obtain completely compressed femtosecond pulse directly from the symmetric/asymmetric FGC output. Here, the "in-house beam-splitting pulse compressor" [23] based on two symmetric FGCs with smoothed laser beams are shown in Figure 9.

With the same optical and laser parameters in the pre-compressor as above single beam design, but higher input pulse energy about 3400 J for the FGC, the input laser energy intensity is calculated to be about 393 mJ/cm$^2$ which is about 1.52 times lower than the grating damage threshold with nanosecond pulse duration. It means that this is a safe input energy intensity because the spatial intensity modulation of the input laser beam has been smoothed to about 1.1 by using the pre-compressor. Considering about 66% efficiency of the FGC due to the loss of the diffraction grating, as high as 1125 J output energy will be achieved at every splitting output. The corresponding peak power is 75 PW for a 15 fs compressed pulse duration. Here, the pulse energy intensity on the G4 or G6 is about 130 mJ/cm$^2$ which is about 1.74 times lower than the grating damage threshold for several tens fs pulse duration. Since the output laser beam is also owing smoothed spatial profile, as a result, about 150 PW laser pulse can be achieved safely with current available optics with tiled-aperture combining of two symmetric MPC outputs.

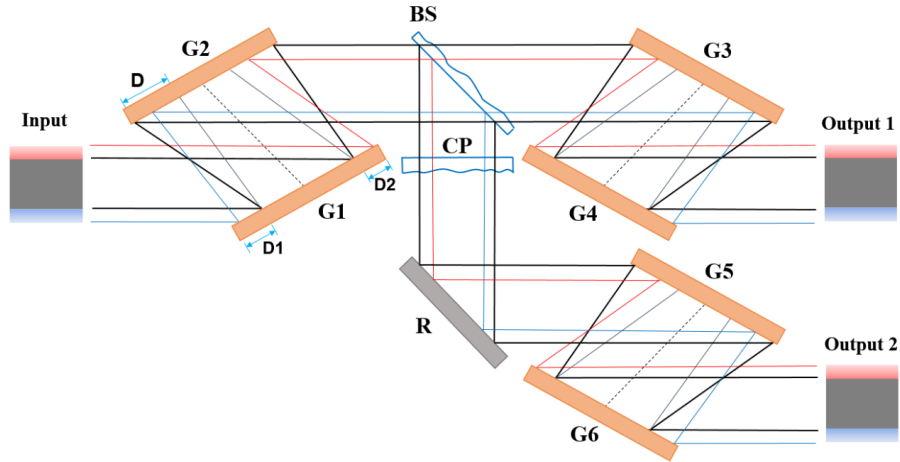

Fig.9. The optical configurations of combining of both the MPC method and the "in-house beam-splitting pulse compressor" method with two symmetric FGCs. G1-G6: diffraction gratings, BS: beam splitter, CP: compensate plate, R: reflective mirror. The pre-compressor and post-compressor are not shown here.

Besides the combining with the "in-house beam-splitting pulse compressor" method, the MPC method can also combine with the tiled-grating method [15]. As we know that the first and the last gratings are easy to be damaged in a FGC system. Then, if the input and output laser beam on the first and the last gratings, respectively, own much larger beam size, it will increases the input and output pulse energy. Previously, this situation has been shown in Figure 5(a), where the first and the last gratings are wider than those of the second and the third gratings. Even though there is no grating with such big size, it is possible to combine the tiled-grating method here. The first grating can be made of two tiled gratings, G1 and G1', where the grating G1 owns the same size of G2 or G3 and the other G1' grating owns a relative narrow wide. The same for the last grating which is made of G4 and G4'. Note that the gap g shown in expression (3) has to be satisfied carefully in this case, which can be improved by changing the grating line density and the incident angle. The optical configurations of combining of both the MPC method and the tiled-grating method with two symmetric FGCs is shown in Figure 10 as following. Together with the modified optical design shown in Figure 8, higher peak power is expected to be achieved. In the case, the laser beam before pre-compressor can own suitable $d/2$ spatial dispersion width, and then the laser beams on all four gratings own smoothed beam profiles. Since the induced spatial dispersion width $D$ could be as wide as 400-500 mm in 10s to 100s PW lasers, it also improves the input and output pulse energy.

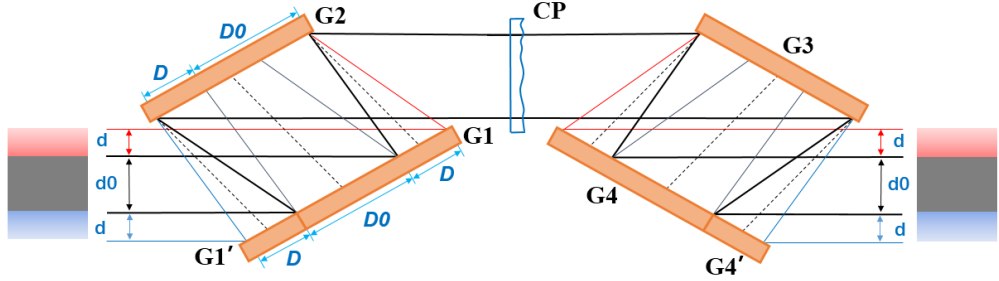

Fig.10. The optical configurations of combining of both the MPC method and the tiled-grating method with two symmetric FGCs. G1-G4: big diffraction gratings with same size, G1', G4': small diffraction gratings with same size CP: compensate plate. The pre-compressor and post-compressor are not shown here.

Note that the laser parameters and the optics parameters in all above designs are not the optimized values. We only show the capabilities of this new MPC method for 10s to 100s PW compression. As we mentioned previously, it is possible further increase the input and output pulse energy by cutting an appropriate amount of the weak edge light owning narrow bandwidth. With negatively chirped hundreds femtosecond or picosecond output from FGC, it may further increase the output pulse energy. Moreover, the output pulse energy can be further improved with homemade extremely large diffraction gratings that can support a size of 500mm ×1000mm input and output laser beam in the near future. It means that several 100s PW laser pulse is expected to be obtained by using the MPC method with extremely large diffraction gratings.

## 7. Discussion

So far, pulse compressor based on diffraction gratings is indispensable for a PW laser facility, where typical FGC is the main compressor. To achieve 10s to 100s PW laser with kJ high energy, multi-beam tiled-aperture combing was the main proposed method in previous reports. There is no usable grating support 100s PW with single typical FGC up to now, which should owe large size, high damage threshold, and broadband diffraction efficiency simultaneously. With three stages of compressors, the proposed novel MPC method is able to achieve 100s PW using single beam. Furthermore, the spatiotemporal properties can be modified conveniently through three stages of compressors which can induce special focal light field. In all, the feasible novel MPC method shows the following exciting properties and obvious advantages:

1). With this MPC design, it is possible to achieve 10s to 100s PW laser without neither beam splitting nor beam combining. As a result, it will improve the output laser stability on many aspects such as beam pointing, pulse duration, and focal intensity, etc. It will avoid complicated and sensitive beam combing process. Furthermore, this design will save hundreds million RMB because it saves many expensive gratings, related optics with meter size, big vacuum chamber systems, and other related optical holders, etc, in comparison to previous 100 PW proposals [7,8,10]. In all, the new setup is simpler, less cost and easier to run.

2). The pre-compressor will induce spatial dispersion to the input laser beam, and then it is convenient to add spatiotemporal modulation optics between the pre-compressor and the main compressor. These added optics will further shape the spatiotemporal properties of the laser beam precisely. For example, it is possible to add a deformable mirror to precisely compensate the spatiotemporal errors induced by the stretcher, the second- and third- gratings in FGC so as to achieve a perfect focal intensity. Otherwise it will induce spatiotemporal aberration that will decrease the final focal intensity [28-30]. The temporal contrast can also be affected by the spatiotemporal aberration which can be measured by using single-shot fourth-order auto-correlator [31]. In comparison to the using of a fixed compensating plate inside the FGC previously [23], this deformable mirror used outside the FGC is programmable, flexible and convenient. It is convenient to handle the wavefront changes and achieve high precise spatial phase compensation. Furthermore, the prism pair based pre-compressor can be used to precisely tune the spectral dispersion of the output laser pulse, which is important for a 15 fs ultrashort laser pulse with a broadband spectrum.

3). The smoothed beam with extremely low spatial intensity modulation will also improve the safety of every FGC in all previous reported PW systems [2]. With smoothed beam, it will avoid the damage of the first and last grating from laser beam with high spatial intensity modulation which mostly coming from the pump beam or diffraction from dust and optical defects before the FGC.

4). Note that, when the spatial dispersion of the input beam is set to opposite to the induced spatial dispersion by the FGC, the input laser beam will not be diffracted out of the second grating G2 if the induced spatial dispersion from pre-compressor is smaller than that induced by the grating pair. Furthermore, the input laser beam can even larger than the output beam from the first grating pair without spectral cutting and energy loss. As a result, the first grating G1 with the maximum size should be considered together with the tiled-grating method, which will also increase the maximum input pulse energy.

5). Completely compressed femtosecond laser pulses with no spatial dispersion but the same spatial intensity modulation as the input beam can be obtained directly at the output. As for obtained smoothed laser beam, of which the spectral dispersion is completely compensated, the post-compressor is simple with the spatiotemporal focusing method. The spatial dispersion on both sides of the beam can be automatically compensated at the focal point by simply using the spatiotemporal focusing effect. Different from traditional focusing, spatiotemporal focusing may induce special focal light field around the focal point.

6). The smoothed output laser beam and negatively chirped pulse will also help to run an additional post-compression process using a thin glass plate. In comparison to recent works [21-22], the smoothed laser beam will avoid the SSSF induced by the high spatial intensity modulation in high spatial frequency. The negatively chirped input pulse based self-compression will avoid the using of meter-sized chirped mirrors for dispersive compensation, which is expensive and not available currently [26]. Furthermore, the thin glass plate is located after the parabolic mirror, there is no optics damage risk in this self-compression process. As a result, even 100s PW laser with few cycle pulse duration is expected to be achieved with nice spatiotemporal properties in the future, which will further increase the focal intensity.

## 8. Conclusion

Pulse compressor became new limitation on achieve extremely high peak power laser pulse with 10s to 100s PW recently. In the paper, a feasible novel MPC method or design is proposed to achieve high-energy 10s to 100s PW lasers which breaks the biggest limitation in this laser system currently. With current available optics, about 100 PW output is expected to be obtained with single beam by using MPC, while more than 100 PW laser pulse can be obtained using single beam together with the tiled-grating method of the first and last gratings. Moreover, 150 PW can be achieved safely by tiled-aperture combining two beams together with the "in-house beam-splitting pulse compressor" method. In comparison to the previous designs by using ≥4 multi-beam tiled-aperture combining method, this new method can simplify the combining beams into two, even single beam without complicated combining process is possible. The key points of the MPC method are using spatiotemporal modified laser as the input and output for the main compressor FGC, which can increase the maximum input and output pulse energies of FGC. An additional pre-compressor is added before FGC compressor to induce the needed beam smoothing, while an additional post-compressor is added after FGC to compensate the output beam in both time domain and space domain. Moreover, during the MPC method, the opposite spatial dispersion between the pre-compressor and the first grating pair of FGC allows even larger grating of G1 and G4 over G2 and G3, which help the success of this method.

By using this MPC method, smoothed and enlarged input laser beam will eliminate the damage risk of both the first and the last diffraction gratings in all PW laser systems [2]. Moreover, together with self-compression using negatively chirped pulse and thin plate in the post-compression stage, even above 100 PW peak power with few cycle pulse duration is expected to be obtained using this MPC in the future. Furthermore, together with the upgrade gratings, the multi-beam tiled-aperture combining method, the tiled-grating method, or negatively chirped pulse based self-compression, this MPC is able to boost the laser peak power to exawatt level in the future, which allows us to explore more frontier fundamental researches.

**Acknowledgments**

The authors would like to thank Dr. Chenqiang Zhao, Dr. Cheng Wang, and Dr. Yi Xu for the discussion on gratings and optics.

**Funding**

This work is supported by the National Natural Science Foundation of China (NSFC) (61527821, 61905257, U1930115), and the Strategic Priority Research Program (XDB16) of the Chinese Academy of Sciences (CAS), and Shanghai Municipal Science and Technology Major Project (2017SHZDZX02).

**Disclosures**

The authors declare no conflicts of interest.